\documentclass[prc,twocolumn,showpacs,superscriptaddress]{revtex4-1}
\usepackage{dsfont}
\usepackage{graphicx,amsmath}
\usepackage{xcolor}

\def\be{\begin{eqnarray}}
\def\ee{\end{eqnarray}}
\def\bc{\begin{center}}
\def\ec{\end{center}}

\newcommand{\lsim}{\stackrel{\scriptstyle <}{\phantom{}_{\sim}}}
\newcommand{\gsim}{\stackrel{\scriptstyle >}{\phantom{}_{\sim}}}

\begin{document}
\title{Examination of the possibility of  $\pi^0$
condensation and  magnetization in freely interpenetrating nuclei
}
\author{D. N. Voskresensky}
   \affiliation{ Joint Institute for Nuclear Research,
		Joliot-Curie street 6,
		141980 Dubna, Russia}
 \affiliation{ National Research Nuclear
    University (MEPhI), 115409 Moscow, Russia}
\begin{abstract}
Conditions are found, at which in    nuclear matter there may appear a spatially nonuniform $p$ wave $\pi^0$ condensate  supplemented by a   spatially varying spontaneous magnetization.  The pion-nucleon interaction and the anomaly contributions to  magnetization are taken into account.  Response of the system on  external magnetic field is also considered. Then  the model of nonoverlapped nucleon Fermi spheres is employed. Arguments are given in favor of  possibility of the occurrence of  the $\pi^0$-condensation and a  spatially varying  magnetization as well as effects of pronounced  anisotropic pion fluctuations at finite pion momentum  in peripheral heavy-ion collisions.  Relevant effects such as response on the rotation, charged pion condensation and other
are  discussed.
\end{abstract}
\date{\today}
 \maketitle

\section{Introduction}
Possibility of appearance of the  $p$ wave pion   condensates in a  dense nuclear matter is analyzed  starting from the beginning of 1970-th, cf. Refs. \cite{Migdal:1971cu,Migdal:1972,Scalapino:1972fu,Sawyer:1972cq,Migdal:1973PL,
Migdal:1973zm,Baym:1973zk,Migdal:1974jn}.    Description of the pion condensation within the chiral $\sigma$ model was suggested in \cite{Campbell:1974qt,Campbell:1974qu,Baym:1975tm,Dautry,Voskresensky:1978cb,
Tatsumi1980,Voskresensky:1982vd}. Various analyses demonstrated that this crystal-like (or liquid-crystal-like) phase of the pion condensation may occur only at the baryon density $n>n_{c}^\pi>n_0$ in the nuclear matter at $N\approx Z$, where $N$ and $Z$ are  the neutron and proton numbers, $n_0=0.16\,$fm$^{-3}\approx 0.48 m^3_\pi$ is the nuclear saturation density, $m_\pi\approx 140$ MeV is the pion mass, $\hbar=c=1$.
After a while focus was shifted to  discussion of effects of pion condensation in neutron stars. Different structures were studied and it was   argued that  energetically favorable  is probably the so-called alternating layer structure of the $\pi^{\pm}\pi^0$ condensates, cf. \cite{Tamagaki76,Tamagaki77,Tatsumi1980,Tamagaki93}.   Also, the pion degree of freedom was considered at a finite temperature with  applications to heavy-ion collisions and neutron stars, particularly to description of their cooling and $r$ modes, cf. \cite{Voskresensky:1978cb,Voskresensky:1982vd,Dyugaev:1982gf,Dyugaev1983,Schulz:1984cb,
Voskresensky:1986af,Voskresensky:1987hm,Senatorov:1989cg,Voskresensky:2001fd,
Grigorian:2016leu,Kolomeitsev:2014gfa}. Effects of fluctuations of the pion field at zero and finite temperature were analyzed in
\cite{Dyugaev:1975dk,Voskresensky:1981zd,Voskresensky:1982vd,Dyugaev1983,Schulz:1984cb}.
 Then the  pion degree of freedom in nonequilibrium nuclear matter was studied, cf.  \cite{Voskresensky:1987hm,Senatorov:1989cg,Voskresensky:1993ud,Voskresensky:2001fd,Kolomeitsev:2010pm}. Numerous number of papers were devoted to  pion condensation in  nuclear systems, relevant references can be found in reviews \cite{Ericson:1988gk,Migdal78,MSTV90,Voskresensky2023}. References \cite{Voskresensky:1981zd,Voskresensky:1982vd,Dyugaev1983,Schulz:1984cb,Senatorov:1989cg,Voskresensky:1993ud} demonstrated enhanced effects of fluctuations keeping pion quantum numbers for $n>n_{c1}\approx (0.5- 0.8)n_0$, when the pion Green function acquires a pronounced minimum at finite momentum $k$ for low pion energy $\omega$. This phase can be associated with the liquid, or glass-like, phase of the pion condensation.

  In \cite{Voskresensky:1980nk} it was shown that in the external magnetic field the charged $p$ wave pion   condensate behaves as the superconductor and first  estimate was done for  the value of the magnetic field in heavy-ion collisions. Possibility of a  $\pi^0\sigma$ running wave condensate in  the neutron matter was studied in \cite{Dautry}.   The pion domain wall structures in dense baryon matter possible due to the axial anomaly  were considered in \cite{SonStephanov,Hatsuda,Hashimoto2015,BraunerYamamoto,Yamamoto,Evans:2022hwr,Eto:2023wul}. Comparison of the ferromagnetic phase of the $\pi^0$ condensate appearing due to the axial anomaly term and the alternating layer structure of the pion condensate  demonstrated that the latter state is probably energetically more favorable \cite{Hashimoto2015}.

Rotational frequencies in nuclei usually do not exceed $\Omega\approx 3\times 10^{21}$Hz, cf. \cite{AfanasievNucl}.
Estimates yield angular momenta $L$ of order of $\sqrt{s}Ab/2\leq 10^6\hbar$ in peripheral heavy-ion collisions of Au $+$ Au  at $\sqrt{s} = 200$ GeV, for the impact parameter $b = 10$ fm, where $A$ is the nucleon number of the nucleus  \cite{Chen2015}. The global
polarization of $\Lambda (1116)$  hyperon observed  by the STAR collaboration in noncentral Au-Au collisions  \cite{Adamczyk2017} indicated existence of a vorticity with  rotation frequency  $\Omega\approx (9\pm 1) \times 10^{21}$ Hz $\approx 0.05m_\pi$, $m_\pi\approx 140$\,MeV is the pion mass.    As a response on  relativistic rotation and magnetic field, there may appear the charged pion condensate in the rotation frame \cite{Zahed,Guo,TeryaevZakharov,Voskresensky:2023znr,Voskresensky:2024ivv,Voskresensky:2024vfx}.  The rotating charged  pion condensate may produce vortices  and  a spontaneous magnetization characterized by the own  magnetic field   ${h}_{\rm L} \approx 3\times 10^{-5}  \Omega$, arising owing to the so-called London moment, with ${h}_{\rm L}$ measured in Gauss and $\Omega$ in rad$/$s. Thus one could have $h\sim {h}_{\rm L} \approx 3\times 10^{17}$G for $\Omega\approx  10^{22}$ Hz, if one dealt with the charged pion  condensate, cf. \cite{Voskresensky:2024vfx}. Gradient of the $\pi^0$ condensate  field also interacts with the baryon axial current via the anomaly \cite{SonStephanov}.

Reference \cite{Gyulassy77}  considered possibility of pion instabilities in the center of mass frame in case of two freely penetrating nuclei. In peripheral heavy-ion collisions, in the momentum space the Fermi seas of nucleons do not overlap, at least for a while,  at a sufficiently large value of the  momentum of the projectile nucleus $p_l$.    Reference \cite{Pirner:1994tt} studied pion instabilities in the laboratory frame and demonstrated that the inhomogeneous pion condensate, with the momentum  $k\neq 0, \vec{k}\perp \vec{p}_{l}$, could be observed in peripheral heavy-ion collisions via  peaks in pion production in a  region of nucleon momenta corresponding to rather large rapidity values. Although  pion  instabilities were not experimentally observed, a feature in the pion cross section in appropriate kinematical region was found at GSI, cf. Fig. 7.5 middle, presented for impact parameter $b\approx 7.9$\,fm in \cite{Bottcher},  and  one may, in principle, hope to observe some peculiar effects related to  pion condensation  in other and more precision  experiments.

Another possibility to get freely penetrating nuclei is associated with  so-called refractive rainbow scattering of nuclei occurring  due to the
refraction of the incident wave by a strongly attractive nucleus-nucleus potential, cf. \cite{Khoa2007} and references therein. In this phenomenon at specific conditions  nuclei scatter elastically and with some probability for a while there may appear  an intermediate cold compound nucleus with the density up to $2n_0$ in a central region, characterized in the momentum space by  only partially overlapped   Fermi spheres of nucleons belonging to colliding nuclei.
Thereby,  possibility of formation of the pion condensation in case of  fully or  partially overlapped Fermi spheres of nucleons   could be, in principle,  checked  in the rainbow experiments. A possibility of experimental check of the results of \cite{Pirner:1994tt} was discussed by the A. A. Ogloblin group in the end of 1990-th, cf. \cite{Ogloblin2003}, but experiment was not performed.

Yet another type of pion instabilities is associated with the Cherenkov-like radiation of pions with the momenta $\vec{k}\parallel \vec{p}_l$ in peripheral heavy-ion collisions at some conditions. Already for $n\approx n_0$ the pion spectrum on a complex plane  gets a minimum for $\omega =\omega (k_m)<m_\pi$ at $k=k_m\neq 0$ for a low temperature $T$ and for $p_l k_m>\omega (k_m)$ this minimum can be in principle  occupied by  pions, cf. \cite{Voskresensky:1993uw,Voskresensky:1995vb,Voskresensky2023}.  Similar effects associated with formation of condensates of excitations can manifest in the flows of  $^4$He \cite{Pitaevskii1984},  cold Bose  gases \cite{BaymPethick2012} and other systems \cite{KolVosk2017}. Let us also in passing indicate a possibility of a  manifestation  of the specific peaks in the $K^+K^-K^0\bar{K}^0$ distributions, cf. \cite{Voskr95}.

Long ago Ref.  \cite{Voskresensky:1980nk} showed that noncentral collisions of heavy ions should be characterized by strong magnetic fields $H\sim h_{\rm VA}\sim H_\pi (Ze^6)^{1/3}\approx (10^{17}\mbox{--} 10^{18})G$ for collision energies less or of the order of   several GeV$\times A$, $Z$ is the charge of the fireball, $H_\pi =m_\pi^2/|e|\approx 3.5\cdot 10^{18}$G, $e^2\approx 1/137$,  $c=\hbar=1$. At ultra-relativistic energies $H$  is  increased typically by the $\gamma_{\rm L}=1/\sqrt{1-v^2}$ Lorentz factor, cf. \cite{Skokov:2009qp}, however quantum effects may result in a decrease of this enhancement effect and the maximum value of the magnetic field can be estimated as $h_{max}\sim h_{\rm VA} Z|e|$, cf. \cite{Voskresensky:2024vfx}. Also, large angular momenta, $L\leq 10^5 \mbox{--} 10^6$, are expected to occur in noncentral heavy-ion collisions, cf. \cite{Huang}, leading  to a   magnetization of  baryons and vector mesons via the Barnett effect. Other mechanisms may also lead to a hadron polarization and magnetization, cf.  \cite{Becattini2008,Becattini:2016gvu,Ivanov:2017dff,Teryaev,Kolomeitsev:2018svb,
Ivanov:2019wzg,Voskresensky:2023znr,Voskresensky:2024ivv,BecattiniLisa} and references  therein. First observations  \cite{Adamczyk2017} have measured the $\Lambda$ and $\bar{\Lambda}$ polarizations   in the energy range $\sqrt{s}_{NN} = (7.7\mbox{--}
200$) GeV.

The paper is organized as follows.
In Sec. \ref{sect-simplified}, I introduce three simplified models for  description of pion degree of freedom   in the isospin-symmetric nuclear matter, $N\approx Z$.
Temperature effects, being estimated, will be then disregarded since further I  concentrate on the case of peripheral heavy-ion collisions.
In Sec. \ref{sect-spin}   spin and magnetic moments associated with $\pi^0$ condensate will be evaluated. Besides the $\pi^0$ condensate  contribution to  magnetization, the axial anomaly contributions will be taken into account. Section \ref{sect-two} describes  pion degree of freedom   in the non-equilibrium model of nonoverlapped nucleon Fermi spheres of colliding nuclei.
Then in Sec. \ref{sect-Gibbs}  the Gibbs free energy density of the $\pi^0$ condensate in presence of the own magnetization and the external magnetic field is constructed and minimized. First, case  $m_\pi\to 0$  and then  realistic case of $m_\pi\neq 0$ will be considered.  In Sec. \ref{sect-estimations} numerical evaluations of critical densities are performed.  Section \ref{sect-relevant} discusses other relevant effects such as rotation, charged pion condensation, chiral-wave condensation, and other.
Some details of calculation of the pion self-energy are deferred to Appendices \ref{Appendix1} and \ref{Amplitude}.

\section{Model for pion self-energy in nuclear matter}\label{sect-simplified}
\begin{figure}\centering
\includegraphics[width=6.8cm,clip]{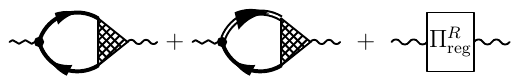}
\caption{Retarded pion self-energy. Bold solid lines symbolize  nucleon and nucleon-hole Green functions, bold double line shows $\Delta$ isobar Green function, hatched vertices include $NN$ and $N\Delta$ correlations,  $\Pi_{\rm reg}^R$ describes a regular residual interaction.
 }\label{PionSelfenergy}
\end{figure}

 The retarded pion self-energy, see  Fig. \ref{PionSelfenergy},  consists of three main  terms, the pole nucleon particle-hole part, $\Pi_P^R$ (the first diagram in Fig. \ref{PionSelfenergy}), the $\Delta$ isobar-nucleon hole term  $\Pi_\Delta^R$ (the second diagram in Fig. \ref{PionSelfenergy}), and a residual regular part, $\Pi_{\rm reg}^R$ (symbolically shown by the third diagram), cf. \cite{Migdal78,MSTV90,Voskresensky:1993ud}. The bold solid lines symbolize  the dressed nucleon and nucleon hole  quasiparticle Green functions, the double bold line is associated with the $\Delta$ isobar quasiparticle Green function.  The wavy line is associated with the dressed pion Green function, with the free pion propagator. Hatched vertices include correlation processes. Within the Fermi-liquid Landau-Migdal theory they are reduced to  taking into account of the nucleon-nucleon and the $\Delta$--nucleon hole  correlations, see respectively Eqs. (\ref{gprimeNN}) and (\ref{DeltaNcor}) below. The retarded pion self-energy enters the Dyson equation for the pion, which solution describes the pion spectrum in  nuclear matter.

I will be interested in description of approximately isospin-symmetric nuclear matter, when  the number of neutrons is approximately equal to number of protons,  $N\approx Z$.
For the purpose  of my exploratory study  it will be sufficient to use simplified expression for the pion self-energy. The real part of the nucleon particle-hole term  of the pion self-energy, see first diagram in Fig. \ref{PionSelfenergy}, renders, cf. \cite{Migdal78,MSTV90,Voskresensky:1993ud},
\be
\Re\Pi^R_P(\omega,k)\approx -\frac{2f_{\pi N}^2  m_N^* (k^2-\omega^2) p_{{\rm F}} \Gamma (g^{\prime},n)\Phi (\omega,k)}{\pi^2}\,,\label{self-enP}
\ee
where $\omega, k$ is the pion 4-momentum, $f_{\pi N}\approx m^{-1}_\pi$ is the $\pi NN$ coupling constant,   $m^*_N(n)$ is the  effective nucleon Landau mass, which  will be parameterized with the help of Eq. (\ref{mef}) of Appendix \ref{Appendix1}, $p_{{\rm F}}(n)$ is the nucleon Fermi momentum, $\Phi(\omega,k,n)$ is the  Lindhard function given by Eq. (\ref{Lindh}) in Appendix \ref{Appendix1}.
For $N=Z$, $p_{{\rm F}}$ is  related to the nucleon density $n$ as $n=2p_{{\rm F}}^3/(3\pi^2)$. The  factor $\Gamma (g'(n),n)$, appeared due to the  dressed vertex in the first diagram of  Fig. \ref{PionSelfenergy},   takes into account the nucleon-nucleon, $NN$, correlations,    cf. \cite{Migdal78,MSTV90},
 \be
\Gamma(g',n)\approx \frac{1}{1+2g'(n)\frac{m^*_N(n)}{m_N} x^{1/3}\Phi (\omega,k,n)}\,,\label{gprimeNN}
\ee
where  $x=n/n_0$,  $g'(n_0)$ is the spin-isospin Landau-Migdal parameter, varying in different approaches employing different normalizations, in  the interval $g'(n_0)\approx 0.6\mbox{--} 1.1$, cf. \cite{Migdal78,MSTV90,Ericson:1988gk}.
The vacuum value of $g'$ is $\approx 1/3$, cf. \cite{Ericson:1988gk},   $g'(n)$, which  will be employed,  is given by Eq. (\ref{gprime}) of Appendix \ref{Appendix1}.

For $0<\omega <kp_{\rm F}/m_N^*$, the imaginary part of $\Pi^R_P$ is as follows \cite{MSTV90,Voskresensky:1993ud},
\be \Im \Pi^R_P(\omega,k)\approx -\beta \omega\,,\quad \beta \approx \frac{f^2_{\pi N}m_N^{*\,2}(n)\Gamma^2 (g',n)k}{\pi}\,.\label{beta}\ee

 In the $\omega, k$ region of our interest  a simplified expression for the $\Delta$ isobar--nucleon hole part of the pion self-energy, shown by the second diagram in Fig. \ref{PionSelfenergy}, is as follows, cf.  \cite{MSTV90,Voskresensky:1993ud},
\be
\Re\Pi^R_\Delta(\omega, k)\approx -\frac{B_\Delta {\omega}_\Delta
\Gamma^2_{\Delta} \Gamma (g'_{\Delta N},\omega)k^2 x}{{\omega}_\Delta^2-\omega^2}\,,\label{pionDelta}
\ee
${\omega}_\Delta(k=0)\approx 2.1 m_\pi$, factor $\Gamma_{\Delta}^2\approx 1/ (1+\beta_0 k^2/m_\pi^2)$ takes into account higher-lying resonances, empirical parameter $\beta_0\approx 0.23$, the $\Delta$--nucleon hole correlation factor (see hatched vertex in the second diagram of Fig. \ref{PionSelfenergy}) is estimated as
\be\Gamma {{(g'_{\Delta }}},\omega)\approx \left[1+\frac{g'_{\Delta}B_\Delta \Gamma^2_{\Delta}\omega_\Delta x}{{\omega}_\Delta^2-\omega^2}\right]^{-1}\,,
\label{DeltaNcor}\ee
empirical value $g'_{\Delta }\approx 0.2$, cf. \cite{Troitsky1981,MSTV90,Voskresensky:1993ud,Suzuki1999}. The coefficient  $B_\Delta\approx 4f^2_{\pi N\Delta} \xi_0 m^3_\pi/9$, the experimental value of the $\pi N\Delta$ coupling is $f_{\pi N\Delta}\approx 2.1 m^{-1}_\pi$, $\xi_0$ is a coefficient, which value is discussed below.

The remaining residual regular part of the pion self-energy, $\Pi^R_{\rm reg}$, shown symbolically by the third diagram in Fig. \ref{PionSelfenergy}, is a smooth but essentially  model dependent function of $\omega, k$. Already in the gas approximation the model dependence appears in dependence on the choice of the going off the mass-shell. This fact was analysed in detail in \cite{Voskresensky:2022gts}. The choice of $\Re\Pi^R_{\rm reg}$ is  discussed in Appendix \ref{Appendix1}, cf. Eqs. (\ref{reg}),  (\ref{Cregcor}), (\ref{reg-on}). The resulting expressions for $\Re\Pi^R_{\rm reg}$ are expressed through   two parameters $C_{\rm reg}$ and $c_{\rm reg}$, which values are constrained by necessity to properly describe pion atoms and to reproduce a reasonable  value of the nucleon $\Sigma$ term.
 The latter quantity can be varied  in broad limits, from 30 MeV up to 60 MeV, cf. a discussion in \cite{Friedman2019,Ikeno}. Recent work \cite{Alexandrou} using lattice calculations  presented the value $\Sigma\approx 41.9\pm 8.1$ MeV.

Further in numerical evaluations I  will use the value $\Sigma \approx 45$ MeV for the nucleon  $\Sigma$ term.  The Model I   employs  experimental value $f_{\pi N\Delta}\approx 2.1 m^{-1}_\pi$ and $\xi_0=1$, which lead to $B\approx 2 m_\pi$. The regular part of the pion self-energy, $\Re\Pi^R_{\rm reg}$, is added according to Eqs. (\ref{reg}), (\ref{Cregcor}) of Appendix \ref{Appendix1}  satisfying the low energy theorems, additionally to \cite{MSTV90} beyond  the gas approximation.  Parameters in Eqs. (\ref{reg}), (\ref{Cregcor}) are chosen to be: $C_{\rm reg}=0.5/m^3_\pi$ and $c_{\rm reg}= 0.5$. The value of the Landau-Migdal parameter $g'\approx 0.7$ is fitted from analysis of the low energy domain ($\omega\ll m_\pi$, $k< 2p_{\rm F}$). The regular part of the self-energy yields extra repulsion for $\omega\ll m_\pi$, $k\neq 0$.

On the other hand, within the chiral-symmetrical model, cf. \cite{Baym:1975tm}, one uses  $f_{\pi N\Delta}\approx 1.7 m^{-1}_\pi$. Thereby in the Model II, I employ    $\xi_0\approx 0.66$ and $f_{\pi N\Delta}\approx 2.1 m^{-1}_\pi$ taking into account that then $f_{\pi N\Delta}\sqrt{\xi_0}\approx 1.7 m^{-1}_\pi$, and I put  $\Re\Pi^R_{\rm reg}=0$. Then the best description of the  low energy domain corresponds  to the choice of the Landau-Migdal parameter  $g'\approx 1$. These modifications compared to the Model I appeared due to  differences in    parameterizations of the regular term  $\Re\Pi_{\rm reg}$.

Description of pion atoms requires essential modification of the pion dispersion law for $\omega$ near $m_\pi$ and $k< m_\pi$ already for $n\leq n_0$. At such $\omega, k$ the pole contribution, $\Re\Pi^R_P$,  is essentially suppressed, as it follows from Eq. (\ref{self-enP}) and Eq. (\ref{Lindh-high}) of Appendix \ref{Appendix1}. With the help of Eq.   (\ref{pionDelta}) and Eqs. (\ref{reg}),  (\ref{Cregcor}), Refs. \cite{Troitsky1981,MSTV90} employing fit of the pion atom  data existed to that time   extracted the pion  branch $\omega^2(k)\approx m^2_\pi+ \widetilde{b} k^2$  with $\widetilde{b}\approx 0.4$ for $\omega \approx m_\pi$, $k\ll m_\pi$  at $n\approx n_0$. Within the Model I employing Eqs. (\ref{self-enP}), (\ref{pionDelta}) and  (\ref{reg}), (\ref{Cregcor})  for $k\to 0$  it will be obtained the value $\widetilde{b}\approx 0.3$.  Within the Model II, employing $\Re\Pi^R_{\rm reg}=0$ and $\xi_0\approx 0.66$ one  reproduces a bit smaller   value, $\widetilde{b}\approx 0.2$. Detailed   analysis of the pion atom data performed basing on the old phenomenological Kisslinger  parametrization for the pion self-energy on the threshold, $\omega\to m_\pi$, $k\ll m_\pi$, cf. \cite{FriedmanGal,Friedman2019}, produces  a bit smaller value $\widetilde{b}\approx 0.15\mbox{--} 0.2$.
With the on-shell parametrization (\ref{reg-on}) for $C_{\rm reg}\approx 0.74/m^3_\pi$, $c_{\rm reg}=0$ taken as in \cite{Kolomeitsev:2010pm}, corresponding to the choice $\Sigma =45$ MeV, employing $\xi_0=0.66$, as in the Model II, one arrives at $\widetilde{b}\approx 0.3$. In this case for $n>n_{c,s}\approx (1.4\mbox{--} 2.5) n_0$ there may occur the $s$ wave pion condensation, cf. \cite{Voskresensky:2022gts}. However, as it was indicated in \cite{Voskresensky:2022gts}, the latter statement could be changed, if there existed a repulsive correlation contribution for $\omega\ll m_\pi$. For instance, modification of the expression  (\ref{reg-on}) by the factor (\ref{Cregcor}) at the choice of parameters $C_{\rm reg}\approx 0.5/m^3_\pi$ and $c_{\rm reg}\approx 0.5$
corresponding to the same value  $\Sigma =45$ MeV would result in the shift of the  $s$ wave  condensation point to the value of the density  $n_{c,s}$ of order of $4 n_0$ and in this case already  at a smaller value $n=n_c^\pi<n_{c,s}$ there may arise the  crystal-like  $p$ wave pion condensate.  I  name the parametrization given by Eqs. (\ref{self-enP}), (\ref{pionDelta}), and Eq. (\ref{reg-on}) modified by the factor (\ref{Cregcor}), with $C_{\rm reg}=0.5/ m^{3}_\pi$, $c_{\rm reg}= 0.5$, the Model III. With the same value $\xi_0=0.66$ the Model III yields a larger attraction at $\omega\approx m_\pi$, $k\ll m_\pi$ than that in Model II and also in the former model  $\Re\Pi_{\rm reg}(\omega \ll m_\pi)<0$. Thereby in the Model III one needs to employ a still larger suppression of the $\Pi_\Delta$ and $\Pi_P$ terms than in the Model II in order to obtain appropriate value of the coefficient $\widetilde{b}$ and a similar description of the total pion self-energy, as in Models I and II for low $\omega$ at $k> m_\pi$. With  such  fitted parameters  the Model III yields a smaller value of the critical density  $n_c^\pi$ than in Models I and II. Thereby, trying to be more conservative, below I will focus attention on the Models I and II.

Closeness to the critical point of the crystal-like pion condensation, $n_c^\pi$, is determined by the value of minimum  of the squared of effective pion gap in $k^2$, e.g. cf. \cite{MSTV90,Voskresensky:1993ud},
\be\widetilde{\omega}^2(k^2,n)=m^2_\pi +k^2+\Re\Pi(\omega=0,k,n)\,,\label{omtild1}
\ee
and in the given case $\Pi=\Pi_P+\Pi_\Delta+\Pi_{\rm reg}$. The value $\widetilde{\omega}^2$ gets a minimum at $k=k_0(n)\neq 0$ only for $n>n_{c1}$, which can be  named the critical density of the liquid-like pion condensation, e.g., cf. \cite{Dyugaev1983,Voskresensky:1993ud}. With increasing $n$ the in-medium pion distribution,  as well as cross-sections of the processes involving $NN$ interaction, get an increase.

\section{ Spin and magnetic moment associated with $\pi^0$ condensate}\label{sect-spin}

Recall that magnetic moment of  nonrelativistic nucleon in vacuum is related to the spin as $g_i M_N\vec{s}_i$,  and the magnetic moment density of protons and neutrons, $i=n,p$, is given by
\be\vec{\cal{M}}_i=g_i M_N\vec{s}_i n_i\,,\quad M_N=\frac{e_p}{2m_N}\,,
\ee
 where    $e_p=-e>0$ is the proton charge, $s_i=1/2$, $\vec{s}_i=\vec{\sigma}/2$, $\vec{\sigma}$ are spin Pauli matrices,  Lande factors  $g_p\approx 5.58$ and $g_n \approx -3.85$, $m_N$ is the nucleon mass in vacuum,  $n_p$ and $n_n$ are proton and neutron densities. Here and below $e_p$ will be measured in units $e^2_p=1/137$. For the medium consisted of the fully polarized  protons and neutrons it would be
 \begin{eqnarray}
&\sum_i \vec{\cal{M}}_{i} =\alpha_N \vec{e}^{\,(p)}=\frac{g_p n_p-g_n n_n}{2}{{M}}_N \vec{e}^{\,(p)}
\,,\label{alphaN}
\end{eqnarray}
since it is energetically profitable to orient the neutron  spin antiparallel to the proton spin, $\vec{e}^{\,(p)}=\vec{s}_p/s_p$. Further,  simplifying consideration I will consider the matter with $N\approx Z$, i.e.,  $n_p\approx n_n\approx n/2$.
The kinetic energy density of the   gas  of nucleon quasiparticles  in case, if  nucleons were fully polarized, renders
\cite{Hashimoto2015},
\be E_{\rm kin}^{(\uparrow)}=\frac{ 3^{5/3}\pi^{4/3}n^{5/3}}{10m^*_N}+m_N^* n
-\frac{\zeta n{{M}}_N \vec{e}^{\,(p)}\vec{h}}{4}\,,\nonumber
\ee
where $\zeta =g_p -g_n$,  $\vec{h}=\mbox{curl}\vec{A}$, $\vec{A}$ is the vector potential of the magnetic field.
The  kinetic  energy density of the very weakly polarized   gas of nucleon quasiparticles  up to terms $\propto (M_N h)^2$ becomes
\be E_{\rm kin}=\frac{1}{2^{2/3}}
\frac{3^{5/3}\pi^{4/3}n^{5/3}}{10m^*_N}+m_N^* n+O[(M_N h)^2]\,.\label{Etwo}
\ee
As it is seen, for $h\to 0$  it is  energetically favorable to have the nonpolarized gas.
Further   small correction  terms $\propto (M_N h)^2$ to the nucleon energy (\ref{Etwo}) will be disregarded.

As it will be argued below, it is energetically favorable to produce  static $\pi$ classical field with a rather small but finite condensate momentum. For the pion energies and momenta  $\omega\ll k\ll 2p_{{\rm F}}$,  the potential of the $p$ wave pion-nucleon interaction is as follows \cite{Ericson:1988gk,Migdal78},
\be
U=-f_{\pi N}\tau_j\sigma_l\frac{\partial\phi_j}{\partial x_l}\,,
\ee
where  $\vec{\tau}$  are the isospin Pauli matrices, $j,l=1,2,3$, $\vec{\phi}=(\phi_1,\phi_2,\phi_3)$ is the pion field. Simplifying consideration  let us assume that only $\phi_3$ classical field is nonzero and $T=0$. The averaged neutron and proton densities in presence of the potential $U$  are given by \cite{Migdal78},
\be
n_i  =\frac{p_{{\rm F}i}^3}{3\pi^2}=\frac{(2m_N^*)^{3/2}}{3\pi^2}\frac{1}{2} \mbox{Tr}_\sigma (\widetilde{\epsilon}_{{\rm F}i}-U_{3i})^{3/2}\,,\label{n-i}
\ee
\be U_{3i}=\mp f_{\pi N}\sigma_j\frac{\partial\phi_3}{\partial x_j}\,,
\ee
the upper sign  is for protons and lower sign is for neutrons, this difference  appeared  due to $\tau_3$ matrix. Trace is performed over spins,
the shifted Fermi energy $\widetilde{\epsilon}_{{\rm F}i}$ is determined from Eq. (\ref{n-i}).

Averaged contribution of the pion-nucleon interaction to the spin density of protons and neutrons is
\begin{eqnarray}
\vec{S}_i^{\,\pi N}=\frac{(2m_N^*)^{3/2}}{3\pi^2}\frac{1}{2} \mbox{Tr}_\sigma [\vec{s}_i (\widetilde{\epsilon}_{{\rm F}i}-U_{3i})^{3/2}]\,.\end{eqnarray}

Simplifying consideration further let us assume that $\phi_3$ and $U_i$ are small and let us retain only linear terms in expansion in  $U_i$. Then $\widetilde{\epsilon}_{{\rm F}i}\approx \frac{p_{{\rm F}i}^2}{2m_N^*}$ and the $z$-component of the averaged spin is
\begin{eqnarray}
({S}_{i}^{\pi N})_z
\approx \pm \frac{f_{\pi N} m_N^* p_{{\rm F}i}\Gamma(g'(n))}{2\pi^2}\frac{\partial \phi_3}{\partial z}
\,.\end{eqnarray}

 Thus the averaged contribution  of the pion-nucleon interaction to the z-component of the density of the magnetic moment  of protons/neutrons is given by
$\vec{\cal{M}}_{i}^{\pi N}=g_i M_N \vec{S}_{i}^{\pi N}\,.$
In spite of that for the isospin-symmetric matter the total spin polarization $(s_{3n}^{\pi N})_z+(s_{3p}^{\pi N})_z\to 0$, there exists significant contribution to the net magnetic moment density of nucleons
\begin{eqnarray}
&\vec{\cal{M}}^{\pi N} =[g_p \vec{S}_{p}^{\pi N}+g_n\vec{S}_{n}^{\pi N}]M_N\approx \alpha_h^{\rm med}\nabla\phi_3\,,\label{pion-nucl-mag-mom}
\end{eqnarray}
\be
 \alpha_h^{\rm med}=\frac{\zeta f_{\pi N}M_N  m_N^*(n) p_{{\rm F}}(n)\Gamma(g'(n))}{2\pi^2}>0\,,\label{alpha-pin}
\ee
where there appeared amplification factor $\zeta=g_p -g_n\approx 9.43$.
Thus the nucleon liquid should have a  strong  response on the magnetic field.

 Let us note that within the $\sigma$ model,  the contribution  of the chiral-wave condensate, $\sigma\pi^0$, to the nucleon spin polarization   was previously considered in \cite{Tatsumi2000}. Let us also notice that in case of the fully polarized matter additionally to the term (\ref{pion-nucl-mag-mom}) there would be yet the purely nucleon contribution (\ref{alphaN}) not associated with the pion condensate.

Another contribution to the magnetic moment of the nucleon may arrive from the so-called  Wess-Zumino-Witten (WZW) axial anomaly term describing
the anomalous interaction of the neutral pion field with the external electromagnetic
field, and a related pion contribution to the baryon current. For example, the WZW term
describes the anomalous $\pi^0\to 2\gamma$ decay. Employing the fields $\Sigma =e^{i\tau_3\phi_3}/f_\pi$, and the nucleon field $A^\nu_N =(\mu_N, \vec{0})$,  one has for the contribution to the net nucleon magnetic moment density, \cite{SonStephanov},
\be \vec{\cal{M}}^{\rm WZW}_N =\alpha_h^{\rm WZW}\nabla\phi_3\,,\,\,\, \alpha_h^{\rm WZW}=\frac{e_p\mu_N}{2\pi^{3/2} f_\pi}, \label{WZW}\ee
where $\mu_N(n)$ is the nucleon chemical potential and $f_\pi\approx 92$ MeV is the pion decay constant.

Reference \cite{Hashimoto2015} included the anomaly contribution studying a $\sigma-\pi^0$ chiral-wave condensate in neutron-rich matter. In their case
\be \vec{\cal{M}}^{\rm WZW}_n =\frac{ie_p \mu_{el}(\sigma +i\phi_3)^\dagger \nabla (\sigma +i\phi_3)}{2\pi^{3/2}f_\pi^2}\,,\label{WZWH}
\ee
where they employ that $\mu_{el}=\mu_p -\frac{1}{2}\mu_n\approx \frac{1}{2}m_N^*$. Assuming that $\sigma\approx f_\pi$, $\phi_3\ll f_\pi$ and replacing $\mu_{el}\to \mu_N$ one would recover Eq. (\ref{WZW}).

Recall that it is  considered the case of weakly polarized matter, cf. Eq. (\ref{Etwo}) in case of the ideal nucleon gas at $N=Z$.
The total ``magnetic'' contribution  to the effective pion Lagrangian density in this  model is thus given by
\be {\cal{L}}_{h}=\left[ (\alpha_h^{\rm med}+\alpha_h^{\rm WZW})\partial_z \phi_3\right]{h}_z, \ee
where  $\vec{h}=\mbox{curl}\vec{A}$ is assumed to be oriented in $z$-direction,  $\vec{A}$ is the vector potential of the magnetic field. The first term is the contribution of the pion-nucleon interaction and second term is the  contribution due to the anomaly, cf. Eqs. (\ref{alphaN}), (\ref{alpha-pin}), and (\ref{WZW}).

Also, let us note that the term ${\cal{L}}_{h}$ yields a contribution not only to the magnetic moment but also to the baryon density $\delta n=\partial {\cal{L}}_{h}/\partial \mu_N$, which is however small for values  $ h$ being smaller or of the order of  $m^2_\pi$ of our interest.

\section{Nonoverlapped Fermi spheres}\label{sect-two}
\subsection{Nucleon distributions}
Our key point is that in peripheral
collisions of heavy ions one may deal with rather cold (with temperature $T\lsim 0.3\epsilon_{{\rm F}N}(n)$) and dilute nuclear matter, cf. \cite{Migdal78,MSTV90,Pirner:1994tt}. The value of the opaque density $n_{\rm op}$, at which the nucleon mean free path is $\lambda_N\approx d/2$, where $d$ is the diameter of the overlap area of nuclei, was estimated in \cite{Pirner:1994tt} as  $n_{\rm op}\approx 1.2 n_0$ (for $d=4$ fm, in the impact parameter range $b\approx (1\mbox{--} 1.7)R$, where $R$ is the radius of the incident nucleus). Therefore  let us further focus on  consideration of  the region with the   density  $n< n_{\rm op}$,  where $n$ is the density in the overlapped region of the colliding nuclei and $n/2$ is the local density in each nucleus at the given value of the impact parameter. As it has been  mentioned, the rainbow scattering of nuclei, cf. \cite{Khoa2007},  gives   another possibility to get for a while the system with only partially overlapped nucleon Fermi spheres,  with the total density reaching up to $n\simeq 2n_0$ in this  case.

The momentum distribution of the two colliding nuclei in the region of their spatial overlap  is  given by the sum of the nucleon distributions shifted in the momentum space,
\be f_{\rm tot}=f(\vec{p})+f(\vec{p}+\vec{p}_l +\vec{k})\,,
\ee
provided one may neglect interactions. For $T\to 0$ one has $f(\vec{p})\approx \Theta (p_{\rm F}(n/2)-|\vec{p}|)$, where $\Theta(x)$ is the step-function.
Excitations from one Fermi sphere are not allowed to
overlap in the momentum space with the ground state distribution in the other
Fermi sphere, provided   ${p}_l >2p_{{\rm F}}(n/2)$ for $k\perp \vec{p}_l$. Then  the factor $f(\vec{p})f(\vec{p}+\vec{p}_l +\vec{k})$ vanishes. One has  $p_l>2p_{{\rm F}}(n_0)\approx 3.8 m_\pi$  already at the nonrelativistic collision energy in the laboratory frame,  $p_l^2/(2m_N)>160$ MeV. For a smaller collision energy nucleon Fermi spheres are partially overlapped and effect under consideration  weakens. At
ultrarelativistic energies effects of the Lorentz contraction of colliding nuclei should be included. Since  all expressions, which are employed for the pion self-energy are valid only for nonrelativistic nucleons,  further the focus will be made on  consideration of collision energies $0.2 \leq E/m_N\leq $ several GeV$\times A$.

Occurring interactions between nucleons in the spatially overlapped region will lead to a decrease of the value $p_l$ with time and an increase of the  temperature. However these collisions are rare  since  typical collisional time between the  particle from the incident   beam and the particle from the target beam is rather long,  $\propto p_l$ for
$p_l\gg p_{{\rm F}}$, cf. \cite{Voskresensky:1991uv}. Thereby,   probability of collisions between nucleons in the region of overlapped nuclei at $n\lsim n_0$ is rather suppressed for the case of the colliding beams, being well separated in the momentum space.

\subsection{Pion self-energy}
 To obtain the pole part of the pion self-energy valid in  case of the nonoverlapped  nucleon Fermi spheres  from the pole part of the equilibrium pion self-energy (\ref{self-enP}),  one should perform  replacement, cf. \cite{Pirner:1994tt},
\be &p_{{\rm F}}(n)\Phi (\omega,k,n)\to  p_{{\rm F}}(\frac{1}{2}n)\nonumber\\
&\times[\Phi (\omega,k,\frac{1}{2}n)+\Phi (\omega -\frac{{k}{p}_l\cos\theta}{m^*_N},k,\frac{1}{2}n)]\,,\label{Phishift}\ee
$\Phi (\omega,k,n)$ is the Lindhard function explicitly presented by Eq. (\ref{Lindh}) in Appendix \ref{Appendix1}, $\theta$ is the angle between $\vec{k}$ and $\vec{p}_l$.
The pion condensation in the isospin-symmetrical matter occurs for $\omega =0$. As it can be seen from the low-energy expansion of the Lindhard function given by Eq.
(\ref{Phi-exp}) of Appendix \ref{Appendix1}, the attraction is largest for  $\vec{k}\perp \vec{p}_l$.
In this case for $\vec{p}_l\parallel x$ one may take $\vec{k}\parallel z$ bearing in mind that the magnetic moment associated with the condensate is parallel to the gradient of the pion field, cf. Eqs. (\ref{pion-nucl-mag-mom}) and (\ref{WZW}). So, in case of  nonoverlapped Fermi spheres of nucleons belonging to the projectile and the target  nuclei the resulting  expression for the pole term of the pion self-energy  becomes
\begin{eqnarray}
&\Pi_P^{(2)}(0,k,\theta,n)=\gamma(n)\Pi_P(0,k,\theta,\frac{1}{2}n)\,,\,\label{self-enP2}\\
&\gamma (n)=\frac{2\Gamma (2g' (\frac{1}{2}n),\theta,\frac{1}{2}n)}{\Gamma (g'(\frac{1}{2}n),\frac{1}{2}n)}\,,\nonumber
\end{eqnarray}
where the nucleon-nucleon correlation factor $\Gamma (g'(n),n)$ is given by Eq. (\ref{gprimeNN}).
For $g'=0$, $m_N^{*}=m_N$,  the overall enhancement factor of the pion-nucleon attraction  in (\ref{self-enP2}) compared to (\ref{self-enP}) would be $2p_{{\rm F}}(n/2)/p_{\rm F}(n)\approx 4^{1/3}$  that  would correspond to effectively four times higher density, favoring occurrence of the pion condensation already for $n<  n_0$ in the model of the nonoverlapped  Fermi spheres.
The $\Delta$ isobar and the regular terms of the pion self-energy depend on the total density $n$ rather than on $p_{\rm F}(n/2)$ and thereby these terms remain the same, as for the case of the fully overlapped Fermi spheres. Squared effective pion gap gets the form
\be
\widetilde{\omega}_{(2)}^2(k^2,\theta,n)=m^2_\pi +k^2+\Re\Pi^{(2)}(\omega=0,k,\theta,n)\,,\label{omegatilde2}
\ee
where $\Pi^{(2)}(0,k,\theta,n)=\Pi_P^{(2)}(0,k,\theta,n)+\Pi_\Delta (0,k,n)+\Pi_{\rm reg} (0,k,n)$.

\subsection{Pion mode in extremely dilute matter}
Let us retain in the expression for the pion self-energy    only terms $\propto x^{1/3}$
 dropping terms $\propto x^{1/2}$ and $\propto x^{1/3}$. In this case $g^\prime \to g_0^\prime  =1/3$, cf. \cite{Ericson:1988gk}, $m^*_N\to m_N$, and for $N=Z$, $\omega =0$, $\vec{k}\perp \vec{p}_l$, $k\ll 2 p_{\rm F}$, the squared effective pion gap, $\widetilde{\omega}^2_{(2)}(k^2,\theta=\pi/2,n)$, acquires the form
\begin{eqnarray}
&\widetilde{\omega}^2_{(2)}(k^2,n)\approx m^2_\pi +k^2[1-\alpha_{(2)}^0\Gamma_0(x)x^{1/3}]+O(k^4)\,,\nonumber\\
&\alpha_{(2)}^0 = \frac{2^{5/3}f^2_{\pi N}m_N p_{\rm F0}}{\pi^2}\,,\,\,p_{\rm F}=p_{\rm F0}x^{1/3}\,,\label{tildeomx}
\end{eqnarray}
$\Gamma_0(x)=1/[1+2^{5/3}g_0^\prime x^{1/3}]$, compare it with Eqs. (\ref{self-enP})--(\ref{pionDelta}) and (\ref{gprime})--(\ref{reg}). For $n\geq n_{c1}$ the  effective pion gap  as function of $k$ acquires minimum at $k\neq 0$  for $k\perp \vec{p}_l$. One has
\be n_{c1}=\frac{n_0}{(\alpha_{(2)}^0-2^{5/3}g_0^\prime)^3}\,.\label{tildeomxnc1}
\ee
Numerical values are $\alpha^0_{(2)}\approx 4.14$, and $n_{c1}\approx 0.034 n_0$. Taking into account correction terms $\propto \sqrt{x}$  results in the value $n_{c1}\approx (0.04 \mbox{--} 0.05) n_0$.
Thus pion fluctuation effects at $k\neq 0$ may start to appear in peripheral heavy-ion collisions already for very small densities, $\geq 0.04n_0$, resulting in some observable effects, such as enhancement of  pion distributions, especially for  $\vec{k}\perp \vec{p}_l$, and the cross sections of the  processes involving the $NN$ interaction. Taking into account of the term  $\vec{k}\vec{p}_l\neq 0$
is discussed in Appendix \ref{Amplitude}.

In artificial case (chiral limit) $m_\pi =0$ one would have  $n_{c1}=n_c^\pi$, where $n_c^\pi$ is the critical density for  occurrence of the  crystal-like phase of the $\pi$ condensation. In realistic case, $m_\pi \approx 140$ MeV, the value $n_c^\pi$ proves to be  significantly higher than $n_{c1}$, and  for $n$ of the order of $n_c^\pi$  terms $\propto x^{1/2}$ and $\propto x$ should be taken into account, see discussion in Sec. \ref{sect-estimations}.

Note that after performing replacements $\alpha_{(2)}^0\to \alpha^0=\alpha_{(2)}^0 /2^{2/3}$ and $g_0^\prime\to g_0^\prime/2$, expressions (\ref{tildeomx}), (\ref{tildeomxnc1}) hold also in case of the equilibrium system (for $N=Z$ under consideration). However in this case also the value $n_{c1}$ is essentially increased (3 times provided terms $\propto x^{1/2}$ are neglected)  and approximation, at which dropped terms $\propto x^{1/2}$ and $\propto x$ can be indeed considered as small, does not work properly.

\section{Gibbs free energy}\label{sect-Gibbs}
\subsection{Expansion in low pion momenta}
Let us continue to apply given  consideration to case of   matter produced in peripheral heavy-ion collisions at $N\approx Z$. For $\omega =0$, assuming that   typical pion momentum is $k\ll 2p_{{\rm F}}$,  let us expand the Gibbs free energy density in a small gradient term  $\nabla \phi_3$ up to  second-gradient order. The total Gibbs free energy density contains now two terms,
\be
G=E_{\rm kin}^{(2)}+G_{h,\phi}\,,
\ee
 where
\be E_{\rm kin}^{(2)}=\frac{1}{2^{4/3}}
\frac{3^{5/3}\pi^{4/3}n^{5/3}}{10m^*_N(n/2)}+m_N^*(\frac{n}{2}) n+O[(M_N h)^2].\label{Etwotwo}
\ee
In comparison with (\ref{Etwo}) extra coefficient $1/2^{2/3}$  appeared in the first term, since now one deals with  nonoverlapped Fermi spheres of particles with the
density $n/2$ in each colliding nucleus,
\begin{eqnarray}
&G_{h,\phi}\approx \frac{(1-\alpha_{1})(\nabla\phi_3)^2}{2}+ \frac{\alpha_{2}(\Delta\phi_3)^2}{2}+
\frac{m_\pi^{*2}\phi_3^2}{2}\nonumber\\
&+\frac{\lambda \phi_3^4}{4}-\alpha_h {h}_z\partial_z\phi_3+\frac{({h}_z-{H}_z)^2}{8\pi}
+\frac{{H}_z^2}{8\pi}\,.\label{GibbsFEhphi}
\end{eqnarray}
Here $\alpha_{1}$, $\alpha_{2}$, $\alpha_h$ are coefficients of  expansion,  $m_\pi^{*}$ is the density dependent pion mass, which includes the $s$ wave contribution to the pion self-energy,  $\vec{H}\parallel z$ is the (external)  magnetic field produced by the two colliding nuclei. For simplicity let us assume that  $H$ is the uniform constant  field,   $\vec{h}$ is the own (internal) magnetic field, which can be produced even in absence of the external field $H$, as it will be shown  below,
and let us take into account that it is energetically profitable to orient $\vec{h}$ parallel $\vec{H}$, $\alpha_h=\alpha_h^{\rm med}+ \alpha_h^{\rm WZW}$. The coupling of the pion-pion interaction $\lambda$  in the nuclear medium is a function of the pion 4-momentum $(\omega, k)$ and the baryon density,  $\lambda (\omega,k,n)=\lambda_{\rm vac}+\lambda_{\rm med}$,   $\lambda_{\rm med}(\omega,k,n)>0$, whereas  $\lambda_{\rm vac}(0,0)= -m^2_\pi/(2f_\pi^2)<0$, cf. \cite{Migdal78,MSTV90}. In-medium effects modify the value of the pion-pion interaction. Change of  sign of the effective value of $\lambda(\omega,k,n)$  in some region of parameters may result in change of the order of  phase transition, cf. \cite{Dyugaev:1975dk,Voskresensky:1978cb}.  To simplify consideration  this possibility will be disregarded and effective $\lambda$ will be thought as a positive parameter. The terms $\alpha_{1}(\nabla\phi_3)^2/2$ and $\alpha_{2}(\Delta\phi_3)^2/2$ are determined by the $\pi^0$ self-energy in the nuclear medium, cf. \cite{MSTV90}, $\alpha_{2}>0$.

\subsection{Limit of negligible pion mass}
Let us first consider a formal limit setting  $m_\pi^*=0$, $\lambda =0$ in (\ref{GibbsFEhphi}). Then one may use
\be\phi=a z\label{chiralwave}
\ee
 as a trial function, for $a=const$. This solution is of the so-called pion wall type, cf. \cite{BraunerYamamoto}. Here it is employed   with taking into account of  nucleon polarization effects.
 Then expression for the Gibbs free energy density is simplified as
\begin{eqnarray}
G_{h,\phi}\approx \frac{(1-\alpha_{1})a^2}{2}
-\alpha_h {h}_z a+\frac{({h}_z-{H}_z)^2}{8\pi}
+\frac{{H}_z^2}{8\pi}\,.\label{Ghphi1}
\end{eqnarray}
Minimizations over $h_z=h$ and then over $a$ yield
\be h=H+4\pi \alpha_h a\,,\,\, a =\alpha_h H/\xi \,,\,\,\xi=1- \alpha_{1}-4\pi \alpha_h^2.\label{hphi0}
\ee
 For $H=0$ and $\lambda =0$  at $n\geq n_{c1}$, determined by the condition
\be \xi(n_{c1})= 0\,,\label{xin}
\ee
there appears instability.
 Stability is recovered provided one takes into account the self-interaction term $\lambda >0$.

For $\xi>0$, $H\neq 0$, $\lambda =0$ substitution of  expressions (\ref{hphi0}) back in (\ref{Ghphi1}) yields
\begin{eqnarray}
G_{h,\phi}=\chi \frac{H^2}{8\pi}\,,\quad \chi\simeq 1- \frac{4\pi\alpha_h^2}{\xi}<1\, .\label{Ghphi11}
\end{eqnarray}
As it is seen, for $m_\pi =0$ the $\pi^0$ condensate appears even for arbitrarily small baryon density as response on the external magnetic field $H$.  The quantity $\chi$
has the sense of the magnetic susceptibility associated with the condensate. Recall that I neglected a small  contribution to the magnetic susceptibility $\propto M_N^2$, cf. (\ref{Etwo}).
For $n<n_{c1}^h$, where the value  $n_{c1}^h$ is determined by the condition $\chi (n_{c1}^h)=0$, or $\xi =4\pi \alpha_h^2$, one has $\chi>0$ and the energy of the pion sub-system  remains to be positive.     In the interval of densities   $n_{c1}>n>n_{c1}^h$ for $\lambda =0$ the energy becomes negative.

\subsection{Case of non-zero  pion mass}
For realistic case $m_\pi\neq 0$. In a model with  specific nonlinear interaction,   solution of the form of  chiral soliton lattice in  strong external field $H$ was considered in \cite{SonStephanov,BraunerYamamoto}. Here, I will focus on a different solution remaining even for $H=0$, being similar to those solutions studied previously in \cite{Dautry,MSTV90} in the limit case  $\alpha_h=0$. At the neglect of in-medium effect on the pion but with taking into account of the anomaly, such kind of solution was considered in \cite{Hashimoto2015}. Additionally let us include effects of the pion polarization in nuclear matter, demonstrate the ferromagnetic effect and apply the results to the case of peripheral heavy-ion collisions.

Let us now find  stable solution with $\pi^0$ condensate for $\xi (n)<0$ in case  $\lambda \neq 0$, $m_\pi\neq 0$. Let us choose the trial function in the form of the standing wave
\be
\phi =\phi_0 \mbox{sin} (k z)\,,\label{standingwave}
\ee
where $\phi_0$ and $k$ are constants provided finite size effects are disregarded.
Then (\ref{GibbsFEhphi}) renders
\begin{eqnarray}
&G_{h,\phi}\approx \frac{(1-\alpha_{1})k^2\phi_0^2\mbox{cos}^2 (kz)}{2}+ \frac{( m^{*2}_\pi m^{2}_\pi+{{\alpha_{2}k^4}})\phi_0^2\mbox{sin}^2 (kz)}{2m^{2}_\pi}\nonumber\\
&+\frac{\lambda \phi_0^4\mbox{sin}^4 (kz)}{4}-{h}_z\alpha_h k \phi_0\mbox{cos} (kz) +\frac{({h}_z-{H}_z)^2+{H}_z^2}{8\pi}.\label{GibbsFEhphik0}
\end{eqnarray}
Minimization of this expression in $h_z=h$ yields the value of the own magnetic field
\be h=4\pi \alpha_h k \phi_0 \mbox{cos} (kz)+H\,,\label{ownh}
\ee
which in presence of the $\pi^0$ condensate is non-zero even for $H=0$. Similar  condensate configurations may appear in  ultra-cold atomic and molecular dipolar gases, cf. \cite{Baym2012}.
Substituting the solution (\ref{ownh}) back in (\ref{GibbsFEhphik0}) and performing averaging of the Gibbs free energy density one has
\begin{eqnarray}
&\overline{G}_{h,\phi}=\frac{\int_0^{L_z} G_{h,\phi}dz}{ L_z}\approx \frac{\widetilde{\omega}^2(k^2,n)\phi_0^2}{4}+\frac{3\lambda \phi_0^4}{32}
+\frac{H^2}{8\pi}\,,\label{Gk0phi}
\end{eqnarray}
where $L_x$, $L_y$, $L_z$ are linear sizes of the large size system,
\be \widetilde{\omega}^2(k^2,n)=m^{*2}_\pi+\xi k^2 +{\alpha_{2}k^4}/{m^{2}_\pi}\label{omtild}
\ee
is the  squared  effective pion gap introduced in Eq. (\ref{omtild1}), expanded in small $k^2$ up to $k^4$ term. This quantity was employed in previous works devoted to the inhomogeneous pion condensation, e.g. cf. \cite{MSTV90,Voskresensky2023}, which however did not incorporate a possibility of the formation of the ferromagnetic state characterized by the own magnetic field (\ref{ownh}).

Minimization of (\ref{Gk0phi})
 over $k^2$ yields the value corresponding to the minimum of the energy for $n>n_{c1}$,
\be k_{\rm{min}}^2=k_0^2=-\frac{{\xi}{{m^{2}_\pi}}}{{2\alpha_{2}}}\Theta (-\xi)\,.
\ee
Substituting this expression back in (\ref{Gk0phi}) and minimizing the latter quantity in $\phi_0$ one finds for $n>n_{c}^\pi$,
\be \phi_0^2 =-\frac{4\widetilde{\omega}^2_0}{3\lambda}\Theta (-\widetilde{\omega}^2_0)\,,\quad \widetilde{\omega}^2_0={{m^{*2}_\pi - \frac{m^{2}_\pi{\xi^2}}{{4\alpha_{2}}}}}\,,\label{omzero}
\ee
where quantity $\widetilde{\omega}^2_0=\widetilde{\omega}^2 (k_{0}^2(n))$ determines the critical point  of the crystal-like phase of the inhomogeneous $\pi^0$ condensation, $n_c^\pi$,
\be \widetilde{\omega}^2_0(n_c^\pi)= m^{*2}_\pi- \frac{{{m^{2}_\pi}} \xi^2(n_c^\pi)}{4\alpha_{2}}=0\,,\label{widetildegap}
\ee
i.e.,
\be \xi(n_c^\pi)=1-\alpha_{1}(n_c^\pi)-4\pi \alpha_h^2(n_c^\pi)=-2
{{\sqrt{\alpha_{2}(n_c^\pi)}}}\,, \ee
and for $n>n_c^\pi$,
\be
\overline{G}_{h,\phi}=-\frac{\widetilde{\omega}^4_0(n)}{6\lambda}\Theta [-\widetilde{\omega}^2_0(n)]\,.
 \ee
The value $n_{c1}$ can be  treated as the  critical density for  appearance of the liquid
(glass-like) phase of the  pion condensation. For $n>n_{c1}$ at $T\neq 0$ there arise strong thermal pion fluctuations at $k\neq 0$,  which result in enhancement of the pion distributions and in  increase of  cross sections of  processes involving the $NN$ scattering, cf.  \cite{Voskresensky:1981zd,MSTV90,Voskresensky:1993ud}.
In this work the contribution of  magnetization is added. Presence of the term $\alpha_h\neq 0$ results in a slight decrease of the value of the critical point, see estimates done in Sec. \ref{sect-estimations} below.

  In  the vicinity of the critical point one has $\widetilde{\omega}^2_0(n)\,{{\propto}}\,  n_c^\pi-n$, and thereby  $(d\overline{G}_{h,\phi}/dn)|_{n=n_c^\pi}\neq 0$ and the   condensate appears  for $n>n_c^\pi$ by the second order phase transition in the mean-field approximation. Fluctuation effects will result in the change of the kind of the transition from the second order to the first order. A more detailed  discussion of these effects goes off our  study here.

 \section{Evaluation of parameters}\label{sect-estimations}
 \subsection{Case of fully overlapped  nucleon Fermi spheres}
 The quantity $\alpha_{1}$ consists of the pole, $\Delta$ and regular terms.
  In equilibrium case of the fully overlapped Fermi spheres of nucleons for $N=Z$   at the density $n$, employing Eqs. (\ref{self-enP}), (\ref{Phi-exp}) for the particle-hole contribution, Eq. (\ref{pionDelta}) for $\Delta$ isobar-nucleon hole term  and, in the Model I, using Eqs. (\ref{reg}), (\ref{Cregcor}) for the regular term and in the Model II setting $\Pi_{\rm reg}=0$, one finds
 \begin{eqnarray}
 \alpha_{1}&=&\alpha_{1}^{P}+\alpha_{1}^\Delta {{-\alpha_{1}^{\rm reg}}}\,,\,\,\\
 \alpha_{1}^{P}&\approx& \frac{2f_{\pi N}^2 m^*_N (n)p_{{\rm{F}}}(n)\Gamma (g'(n),k=0)}{\pi^2}\,,\nonumber\\
 \alpha_{1}^\Delta&\approx& \frac{2m_\pi\xi_0 \Gamma(g'_\Delta,n,k=0)x}{{\omega}_\Delta}
 \,, \nonumber
 \end{eqnarray}
 where $x={n}/{n_0}$, and
\begin{eqnarray}
 \alpha_{2}&=&\alpha_{2}^P+\alpha_{2}^\Delta\,,\\
  \alpha_{2}^\Delta&\approx& \frac{2m_\pi \beta_0\xi_0 \Gamma^2(g'_\Delta,n,k=0)x}{{\omega}_\Delta}\,,\nonumber\\
 \alpha_{2}^{P}&\approx& \frac{\alpha_{1}^{P}\Gamma (g'(n),k=0)m^2_\pi}{12p^2_{\rm F}}\,.\nonumber
 \end{eqnarray}
The contribution $\alpha_h^{\rm med}+\alpha_h^{\rm WZW}$ is determined by Eqs. (\ref{alpha-pin}), (\ref{WZW}).  In the Model I with parametrization  (\ref{reg}), (\ref{Cregcor}) of Appendix \ref{Appendix1} one  gets $\alpha_{1}^{\rm reg}=2C_{\rm reg}^*n_0x$
and the effective pion mass $m^{*2}_\pi\approx m^2_\pi(1 +C_{\rm reg}^*n_0x)$.
 Within the Model II the regular term $\alpha_{1}^{\rm reg}=0$ and $m^{*}_\pi=m_\pi$. In the Model III, $\alpha_{1}^{\rm reg}=0$ and $m^{*\,2}_\pi\approx m^2_\pi (1-C_{\rm reg}^*n_0x)$.

 Employing values  $m_\pi \approx 140$ MeV, $m_N\approx 6.7m_\pi$  from (\ref{alpha-pin}), (\ref{WZW}) one obtains
  \begin{eqnarray}
 \alpha_h^{\rm med}\approx 0.04(1-0.2\sqrt{x})x^{1/3}\Gamma(g'(x)) \,,\label{alphahmed}
  \end{eqnarray}
 \begin{eqnarray}
 \alpha_h^{\rm WZW}\approx 0.08(1-0.2\sqrt{x}) \,.\label{alphahWZW}
  \end{eqnarray}

  Employing (\ref{alphahmed}), (\ref{alphahWZW}) one may  estimate  correction to the value $\xi$ in Eq. (\ref{hphi0}),
   which appears due to  presence of the own magnetic field, $\delta_h=-4\pi \alpha_h^2\approx -(0.06\mbox{--} 0.08)$ for densities $n\leq (3\mbox{--}4)n_0$ of our interest. This term favors occurrence of the pion condensation at a bit smaller density, since the value $\widetilde{\omega}_0^2$ decreases, cf.  Eq. (\ref{omzero}). For $N=Z$ and $H=0$  the typical value of the amplitude of the spatially varying own  magnetic field $h$, see Eq. (\ref{ownh}), can be estimated as of the order of $m^2_\pi$, which value is larger than the London field $h_{\rm L}$ estimated in Introduction.

In the assumption of the second-order phase transition, the crystal-like pion condensation  occurs  in the Model II  for $n>n_{c}^\pi$, being estimated as $ 2 n_0$, and in the Model III at $n_{c}^\pi$, being estimated as $1.5 n_0$. In the Model I one  estimates $n_c^\pi$ as ${{3.5}} n_0$. Unfortunately  these quantitative estimates are essentially model dependent. With taking into account   possibility of the first-order phase transition, estimated quantities $n_{c}^\pi$ would be shifted to smaller values.   Moreover, the larger value of $n_{c}^\pi$ is the higher is uncertainty in the estimation since  expansion of the Gibbs free energy is performed only up to terms of the order of $k^4$.

It should be also noticed that already for $n>n_{c1}$ and especially in the vicinity of the critical point $n_{c}^\pi$ one may expect occurrence   of enhanced fluctuations with the pion quantum numbers with  $k\approx k_0\neq 0$. In the Model I the value  $n_{c1}$ is roughly estimated as $0.5 n_0$, in Model II, as $n_0/4$ and in Model III, as $0.2 n_0$.

\subsection{Case of nonoverlapped  Fermi spheres}
As it has been mentioned, in case of the peripheral heavy ion collisions at intermediate collision energies  one, for a while,  feasibly deals with  nonoverlapped Fermi spheres each corresponding to the density $n/2$. Then  the quantities  $\alpha_{1}^\Delta -\alpha_{1}^{\rm reg}$, $\alpha_{2}^\Delta$ and $\alpha_h^{\rm WZW}$ depending only on the density are not changed, whereas values $\alpha_{1}^{P}$, $\alpha_{2}^P$ and $\alpha_h^{\rm med}$ should be replaced by  $\alpha_{1}^{P,2}$, $\alpha_{2}^{P,2}$, $\alpha_h^{\rm med,2}$ respectively:
\begin{eqnarray}
\alpha_{1}^{P,2}(n)&=&\gamma(n) \alpha_{1}^{P}(n/2)\,,\,\nonumber\\
\alpha_{2}^{P,2}(n)&=&\gamma^2(n)\gamma_2\alpha_{2}^{P}(n/2)\,,\nonumber\\
\alpha_{h}^{\rm med,2}(n)&=& \gamma(n)\alpha_h^{\rm med}(n/2)\,,\label{alpha2Fermi}
 \end{eqnarray}
where factor $\gamma(n)$ is determined in Eq. (\ref{self-enP2}).
 For $\vec{k}\vec{p}_l\neq 0$  there appears extra pre-factor
\be \gamma_2=\frac{1+ p_l^2p^{-2}_{\rm F}\cos^2 \theta }{2}\,. \label{gamma2}\ee
To obtain (\ref{alpha2Fermi})  expansion of (\ref{omegatilde2}) in low $k$ was performed taking  into account shift of $\omega$ in the one of the Lindhard functions (\ref{Phishift}) and  expansion of (\ref{Phi-exp}) in small $\omega$ was used.  Equation (\ref{gamma2}) is valid only for $p_l^2p^{-2}_{\rm F}\cos^2 \theta \ll 1$. For $\vec{k}\perp\vec{p}_l$ one has $\gamma_2=1/2$. In opposite limit, $p_l^2p^{-2}_{\rm F}\cos^2 \theta \gg 1$, one of the two Lindhard functions, $\Phi (\vec{k}\vec{p}_l/(kv_{\rm F}), k)$, is strongly suppressed, cf. Eq. (\ref{Lindh-high}),
and  one can approximately put $\alpha_{1}^{P,2}(n)\approx \alpha_{1}^{P}(n/2)$, $\alpha_{2}^{P,2}(n)\approx   \alpha_{2}^{P}(n/2)$, $\alpha^{\rm med,2}_{h}(n)\approx  \alpha^{\rm med}_{h}(n/2)$.  Thus, it is seen that the squared  effective pion gap given by Eq. (\ref{omegatilde2})  shows a sharp dependence on the angle $\theta$ between vectors $\vec{k}$ and $\vec{p}_l$. Such a dependence could be manifested in experimental distributions of pions and other particles undergoing nucleon collisions,  cf. Eq. (\ref{ratio1}) in Appendix \ref{Amplitude}.

For Models I, II and III, factor $\gamma$ proves to be a smooth function of $g'(n)$ and $n$. In the interval of values $g'(n)\approx 0.7\mbox{--} 1$,  $n\approx (0.5\mbox{--} 1)n_0$,  one finds  $\gamma \approx 1.4\mbox{--} 1.6$. Appearance of the correction term $-4\pi \alpha_{h,2}^2\approx -0.1$ in the expression for $\xi$, shows that the magnetic effect favors a moderate decrease of  the  value of the critical density of the pion condensation.
As the result, the crystal-like pion condensation occurs in the Model I for $n_{tot}>n_{c}^\pi\approx {{(1.5\mbox{--} 2)}}n_0$,   in the Model II for $n_{tot}>n_{c}^\pi\approx (0.6\mbox{--} 0.7) n_0$, and in the Model III  for  $n_{c}^\pi\approx 0.5 n_0$. Also, it should be noticed that for the case of nonoverlapped  Fermi spheres the value  $n_{c1}$ is estimated as $0.05 n_0$, i.e., being  much smaller than $n_0$, and thereby in the broad region of densities (and impact parameters) one may expect manifestation of enhanced pion (and may be also antikaon and some other particle) distributions at $k\neq 0$.

\section{Other relevant effects}\label{sect-relevant}
Let us briefly discuss other relevant effects.

\subsection{Response on rotation}\label{sect-rotation}
 Even in case of weakly interacting beams of nuclei the viscosity has  nonzero value, which  causes appearance of the nonzero angular momentum.
The  anomaly  produces the WZW-contribution to the Gibbs free energy density (\ref{GibbsFEhphi})  due to rotation associated with the $\pi^0$ condensate,
\be \delta E_{\omega}^{\rm WZW}=-\alpha_\omega m_N\nabla\phi_0\vec{\omega},\,\, \alpha_\omega=\frac{\mu_N\mu_I}{2\pi^2 f_\pi m_N},\label{anomrot}\ee
where $\mu_I$ is the isospin chemical potential, $\omega$ is the rotation frequency,  cf. \cite{Yamamoto}. The latter quantity can be presented as sum of two terms: $\vec{\omega}=\vec{\Omega}+\vec{\omega}_{\rm own}$, $\vec{\Omega}$ is the external rotation frequency and $\vec{\omega}_{\rm own}$ is the own self-rotation term. The latter term is determined from minimization of the sum  $\delta E_{\omega}^{\rm WZW}+I(\omega-\Omega)^2/2$, where $I$ is the moment of inertia of the system. For  extended systems when $I$ is large, ${\omega}_{\rm own}$ becomes tiny.

Employing Eq. (\ref{anomrot}) one finds
   \begin{eqnarray}
 \alpha_\omega\approx 0.5 (1-0.2\sqrt{x})\mu_I /m_N\,.
  \end{eqnarray}
In case of the isospin-symmetric matter, $\mu_I=0$, the contribution (\ref{anomrot}) vanishes. For a neutron-rich matter $\mu_I>0$ and    $\delta E_{\omega}^{\rm WZW}<0$.  This circumstance can result in    appearance of some  isospin asymmetry, $N\neq Z$, in rotating nuclear matter, e.g., in peripheral heavy-ion collisions.

For $N=Z$ and $x=1$  one estimates $\alpha_h^{\rm med}/\alpha_h^{\rm WZW}\approx 0.2$ and the main contribution to the magnetic part of the energy comes from  the term $\alpha_h^{\rm WZW}$.  With increasing isospin asymmetry, $\alpha_\omega$ begins contribute and the value of the critical density for  appearance of the crystal-like phase of the $\pi^0$ condensation decreases.  Using for estimates values  $\Omega\approx 0.1 m_\pi$ and $h\approx m_\pi^2$ and $N\approx 1.5Z$ one may find that the rotation contribution to the Gibbs energy can become  of the order of that from $\alpha_h^{\rm WZW}$. This circumstance results in a lowering of the values of the critical densities estimated above, e.g.,  in the Model I for the case of nonoverlapped Fermi spheres the value  $n_c^\pi$ can be estimated as $\approx 1.5 n_0$, and  $n_c^\pi\leq 0.5 n_0$ in Models II and III.

 Reference \cite{Buzzegoli} argued  that at initial stage of the heavy-ion collision one may have  $|eH|\gg \Omega^2$ and at a latter stage oppositely  $|eH|\ll \Omega^2$. It should be noted that in presence of the $\pi^0$ condensate the own magnetic field  is such that   up to the  freeze out time  $|eh|\gg \Omega^2$.

At a nonrelativistic rotation the transition from the laboratory $\rm l$-frame to the
$\rm r$-frame rotating together with the medium is given by the relation $E_{\rm r}=E_{\rm l}-\vec{\omega} \vec{J}$, where $J$ is the total momentum (angular momentum and spin). Thus there may appear additional rotation term in the Gibbs free energy density associated with the neutral pion condensate.

\subsection{Charged pion condensation}
At ignorance of electromagnetic effects (for $e\to 0$) at $N=Z$ for $n<\mbox{min}\{ n_{c\pm}^{\pi}, n_{c0}^{\pi}\}$  the $\pi^{+},\pi^{-},\pi^0$ mesons have the same spectrum.  Generally speaking critical densities $n_{c\pm}^{\pi}, n_{c0}^{\pi}$ are not equal each other, since  the phase transition to  inhomogeneous condensate state  $k\neq 0$ is always of the first order owing to effect of the quantum and thermal fluctuations, cf. \cite{Dyugaev:1975dk,Voskresensky:1981zd}. For $T=0$ this difference is rather small and can be neglected with some accuracy, cf. \cite{Migdal78}.
Condensate of $\pi^0$ is described  by  real field (e.g., of the form of the standing wave (\ref{standingwave})), whereas the $\pi^{\pm}$ condensate is described by complex field, e.g. of the form of the running wave $\phi=\phi_0e^{\vec{k}\vec{r}}$, and fluctuation terms are different in mentioned cases. Also, in case of the $\pi^0$ condensation the anomaly attractive WZW term contributes, cf. \cite{BraunerYamamoto}. Thereby there arises question, condensate of which kind is energetically most preferable.
In this work   focus is made  on consideration of the $\pi^0$ condensate. A weak  magnetic field is ordinary repelled from  charged $\pi^{\pm}$ condensate owing to the Meissner effect, cf. \cite{Voskresensky:1980nk}.
However,   as it  has been mentioned, in case of the rotating charged medium, presence of the normal (proton) current causes appearance of a compensating  superconducting $\pi^-$ condensate  current in the rotating frame, which may result in occurrence of the  London moment and the  corresponding London uniform magnetic field
 $\vec{h}_{\rm L} = -(2m_\pi/e_p)\vec{\Omega}\approx 3\cdot 10^{17}$G for $\Omega\approx  10^{22}$ Hz,  $e_p$ is the charge of proton, cf. \cite{Voskresensky:2024vfx}.  This value of $h$ is  smaller than the typical value of the own magnetic field, $h$ of the order of $m_\pi^2$, which may appear in case of the $\pi^0$ condensation. In a more detail these effects will be studied elsewhere.

 \subsection{Chiral wave solution}
  For description of the $\pi^0$ condensate field within the model $\lambda\phi^4$    the trial function in the form of the standing wave (\ref{standingwave}) was used. If one worked within the $\sigma$ model, one could  choose the trial function for the complex field $\phi=\sigma +i\phi_3$ in the form of the running wave $\phi=\phi_0 e^{\vec{k}\vec{r}}$, similar to that takes place in case of the charged pion condensate.  Then the  condition (\ref{WZWH}) would lead to the constant magnetization $\vec{M}_N^{\rm WZW}=e_p\mu_N \vec{k}\phi_0^2 /(2\pi^{3/2}f^2_\pi)$ instead of the spatially varied one given by Eq. (\ref{WZW}). The energy is gained in the former case.
  Reference \cite{Voskresensky:2023znr} has discussed a possibility of the $\sigma\pi^0$ vortex condensate in the rotating nuclear matter, which permits coexistence of  superfluidity and rotation in the neutral $\sigma\pi^0$ sub-system, similarly to the case of superfluidity in rotating He-II.

  In presence of rotation in  superfluid system there exists the critical angular velocity, above which the rotating system produces the lattice of vortices. In a more detail these effects will be considered elsewhere.

\section{Conclusion}
In this work focus was made on  manifestation of the nucleon polarization due to the
$p$ wave in-medium  nucleon interaction with the  $\pi^0$ field.
Description of the $\pi^0$ spectrum was employed for  approximately isospin-symmetric nuclear matter, $N\simeq Z$, at very low temperatures. Three models were formulated, see Appendix \ref{Appendix1}, which differently treat the residual  part of the pion self-energy. The Model I following \cite{MSTV90} takes into account the low energy theorems. The simplest Model II neglects the residual  term $\Pi_{\rm reg}$ in the pion self-energy, as it has been suggested in  \cite{Migdal:1974jn,Baym:1975tm,Voskresensky:1982vd,Ericson:1988gk} and in other early works. The Model III  follows the on-mass-shell treatment \cite{Delorm1992,Erikson1994,Kolomeitsev2003}.   The parameters of the Models I, II and III were fitted to appropriately treat the region of the pion momenta  $\omega\approx m_\pi$, $k\ll m_\pi$. Models I and II also appropriately describe the low pion energy domain, $\omega\ll m_\pi$, $k\geq m_\pi$ at $n$ of the order of $n_0$,   cf. \cite{Migdal78,Ericson:1988gk,MSTV90} and refs therein. The Model III yields a stronger attraction in this region of $\omega,k$. In  Models I, II and III a minimum of  square of the effective pion gap $\widetilde{\omega}^2 (k^2,n)$  at $k=k_0\neq 0$, cf. Eq. (\ref{omtild1}), arises for $n>n_{c1}$, being  estimated as $0.5 n_0$,  $0.25 n_0$ and $0.2 n_0$, respectively. This, glass-like (or liquid) phase of pion condensate,  is characterized by  enhanced,  for $k\neq 0$,  quantum and especially thermal (for $T\neq 0$) fluctuations heaving  pion quantum numbers. The optimal value of $k\neq 0$ increases with increasing density reaching values larger than $m_\pi$ for $n> (0.5\mbox{--} 1)n_0$. Note that in the models describing  pion spectra in inclusive processes at  heavy-ion collision energies $\lsim 1$GeV$\times A$ one usually gets for the break up density the value $n_{\rm br}\approx (0.5\mbox{--} 0.7)n_0$, which could be associated with effect of  suppression of  pion fluctuations with $k\neq 0$ for  smaller $n$, cf. \cite{MSTV90,Voskresensky:1993ud}. It would be also interesting to seek a possible correlation between appearance of pronounced pion fluctuations with $k\neq 0$ for $n>n_{c1}$  and   the $\alpha$ clustering at $n>n_{\rm Mott}$, where $n_{\rm Mott}$ is the critical density for the Mott transition, cf. \cite{Typel:2009sy} and references therein,  as well as with  occurrence of the Pomeranchuk instability in the scalar $NN$ interaction channel, cf. \cite{Kolomeitsev:2016zid}.  Note that at zero temperature a transition at a critical radius from a homogeneous system to a tetrahedral-clustered configuration, corresponds to $n_{\rm Mott}\approx 0.3 n_0$, being of the order of $n_{c1}$, e.g., as indicated by development of the non-axial octupole $\beta_{32}$ deformation in diluted $^{16}$O, cf. \cite{Davies2025}.

In Sec. \ref{sect-spin} the spin and magnetic moment associated with the crystal-like $\pi^0$ condensate were calculated and the axial anomaly Wess-Zumino-Witten contribution was added. It proved to be that for $N\approx Z$ there is no   purely nucleon term in the energy density, being linear in  spin, cf. Eq. (\ref{Etwo}). In spite of that  the averaged spin density $(s_{3n}^{\pi N})_z+(s_{3p}^{\pi N})_z\to 0$ for $N=Z$, in presence of the $p$ wave $\pi^0$ condensate  there appears   linear  contribution to the net magnetic moment proportional to  gradient of the condensate field.  One  term  given by  Eqs. (\ref{pion-nucl-mag-mom}), (\ref{alpha-pin}) proved to be of the first-order in the  $f_{\pi N}$  coupling constant  and  the other one appeared due to the axial anomaly. Numerically the latter term  proved to be several times larger than   the  term $\propto f_{\pi N}$.

Then in Sec. \ref{sect-two}  focus was made   on  possibility of manifestation   of the $\pi^0$ condensation and magnetization in peripheral heavy-ion collisions. For this porpoise the model of
nonoverlapped Fermi spheres of nucleons belonging to projectile and target nuclei with $N\approx Z$  was employed, cf. \cite{Pirner:1994tt}. Probability of collisions between nucleons in the region of spatially overlapped nuclei at $n < n_0$ proves to be rather suppressed for the case of the colliding beams, being well separated in the momentum space. Thereby, the model assumes that  for a while excitations from one nucleon Fermi sphere are  not allowed to overlap in the momentum space with the ground state distribution in the other
Fermi sphere provided   ${p}_l >2p_{{\rm F}}(n/2)$ for $\vec{k}\perp \vec{p}_l$, where $p_l$ is the momentum of the projectile nucleus. It occurs for   heavy-ion collisions at the collision energy in the laboratory frame  $>160$ MeV. For a smaller collision energy nucleon Fermi spheres are  partially overlapped. At ultrarelativistic energies effects of the Lorentz contraction of colliding nuclei should be included.
The pole part of the pion self-energy proves to be $\propto p_{\rm F}\propto n^{1/3}$. Therefore neglecting $NN$ correlation effects, for nonoverlapped for a while nucleon Fermi spheres (for case of the isospin-symmetric nuclei) one could  gain in the pion-nucleon attraction the factor up to $2\times (n/2)^{1/3}$ compared to $n^{1/3}$, that would lead to occurrence of the $p$ wave pion condensation at a smaller density than in  case of the equilibrium system  of nucleons.

In the model of
nonoverlapped Fermi spheres of nucleons the value of the critical density $n_{c1}$ proved to be very small, $n_{c1}\approx (0.04\mbox{--}0.05) n_0$. It is remarkable that this estimate is  only a weakly  model dependent. This may stimulate experimental  search of  effects of  enhanced pion fluctuations at $k\neq 0$, $\vec{k}\perp \vec{p}_l$, in peripheral heavy-ion collisions.  Appendix \ref{Amplitude} demonstrates presence of a sharp angular ($\vec{k}\vec{p}_l$) dependence of the $NN$ amplitude, which enters the cross sections of the pion and other particle production in $NN$ collisions for $n>n_{c1}$.  Such a dependence could be manifested in experimental distributions of pions, and may be kaons and other particles undergoing nucleon collisions. Let us also mention that Ref. \cite{Kolomeitsev:2016zid} indicated a possibility of appearance of a metastable nuclear state at $n\approx (0.05\mbox{--} 0.1) n_0$ owing to a possible condensation of the scalar quanta, which may occur as  result of the Pomeranchuk instability  in the dilute nuclear matter. It would be interesting to study whether growth of pion fluctuations at $k\neq 0$ for $n>n_{c1}$ could stimulate population of the mentioned metastable state in peripheral heavy-ion collisions.

Results of Sec. \ref{sect-Gibbs} hold in both cases of nonoverlapped and overlapped nucleon Fermi spheres.
The $\pi^0$-condensate contribution to the Gibbs free energy density $G_{h,\phi}$ was calculated  in presence of  external uniform magnetic field $\vec{H}\parallel z$ and feasibly the own magnetic field $\vec{h}$.  To proceed analytically  the low $k$-momentum expansion up to the second order in $\nabla\phi$  ws employed. The field $h$ was then found by minimization of $G_{h,\phi}$.

First an artificial case of negligible pion mass was considered. In this case the solution  for the pion field proves to be of the  pion wall type, cf. \cite{BraunerYamamoto}. In absence of the  magnetic field  $H$, the pion instability would arise for $n>n_{c1}$ given by condition (\ref{xin}). Stability is recovered by taking into account  the pion-pion repulsive self-interaction,  in the model $\lambda\phi^4/4$ for $\lambda>0$.
It was shown that in presence of the external magnetic field  $H$ for $\lambda=0$    appearance of the condensate proves to be  energetically favorable even for arbitrary low density $n$, cf. Eq. (\ref{Ghphi11}). In the density interval,  $n_{c1}>n>n_{c1}^h$, the energy of the pion sub-system is negative.
For $m_\pi =0$ at $n>n_{c1}$ stability can be recovered only with taking into account of the pion self-interaction, for  $\lambda >0$.

Then the focus was turned to  the realistic case $m_\pi \approx 140$ MeV. The critical point of the $p$ wave $\pi^0$ condensation in  assumption of the second-order phase transition is determined by  zero of the effective pion gap, cf. condition (\ref{widetildegap}). In case of the $p$ wave $\pi^0$ condensation in absence of the external magnetic field there arises periodic own magnetic field, cf. (\ref{ownh}). Its magnitude is estimated to be of the order of $ m^2_\pi$. In case of peripheral heavy-ion collisions the magnetic field and  rotation  work in favor of occurrence of the $p$ wave $\pi^0$ condensation at a  smaller density. Occurrence of the condensation causes appearance of a self-magnetization and a weak self-rotation. The latter causes a modification of the $N/Z$ ratio compared with the initial value $N/Z$ characterizing colliding nuclei.

Following  numerical evaluations performed in Sec. \ref{sect-estimations}, in case of the equilibrium system characterized by fully overlapped nucleon  Fermi spheres the crystal-like pion condensation, being treated in the mean-field approximation,  does not occur  by the second-order phase transition up to   $(3\mbox{--}3.5) n_0$ in the Model I, but it may occur for $n>n_c^\pi\approx 2 n_0$ in the Model II, and already for $n>n_c^\pi\approx 1.5 n_0$ in the Model III. Self-magnetization and self-rotation of the nuclear system  result in a moderate  decrease of $n_c^\pi$.

 In case of nonoverlapped nucleon Fermi spheres, with taking into account of the magnetic and rotation effects  the value $n_c^\pi$ was estimated to be   $ (1.5\mbox{--} 2) n_0$ in the Model I,  $(0.6\mbox{--} 0.7) n_0$ in the Model II, and $\approx 0.5n_0$ in the Model III.
 Especially it would be interesting to check the experimental data on  presence or absence of the specific anisotropy in pion distributions as function of the angle between $\vec{k}$ and $\vec{p}_l$ for $k\gsim m_\pi$. In passing,  let us also once more recall  that one expects occurrence of densities up to $2n_0$ in the elastic nuclear rainbow scattering.

 Also, possibilities of the $p$ wave $\pi^\pm$ condensation and the $\sigma\pi^0$ chiral wave condensation were discussed.

 Concluding,  evaluations performed in all three Models I, II and III, although being model dependent, can be treated as rather optimistic to seek manifestation of  enhancement of pion fluctuations at $k\neq 0$ for $\vec{k}\perp\vec{p}_l$, a significant specific anisotropy of the pion  distributions, as well as  manifestation of  the $\pi^0$ condensation and a spontaneous magnetization and a self-rotation in peripheral heavy-ion collisions.

 \acknowledgments
I thank Yu. B. Ivanov, E. E. Kolomeitsev and N. Yamamoto  for fruitful discussions.

\appendix
 \section{Details needed for calculation of the pion self-energy}\label{Appendix1}

  Let us allow for a smooth $n$-dependence of $g'$ for $n\lsim n_0$. To be specific we employ the parametrization:
 \be
 g'(n)\approx 0.33+\xi_1\sqrt{x}\,,\quad x=n/n_0\,,\label{gprime}\ee
 for $n\leq n_0$ and $g'=0.33+\xi_1$ for $n>n_0$, which is more smooth than the linear approximation  used previously in \cite{Mig}. In the Model I let us take $\xi_1=0.37$ and in the Model II, $\xi_1=0.67$. Also let us take into account a density dependence of the effective  {{Landau}} nucleon mass,
 \be
{{  m_N^*= \sqrt{m_{D}^{*\,2}+p_{\rm F}^2}\,,\quad m_{D}^*\approx m_N (1-0.2\sqrt{x})\,,}}\label{mef}
 \ee
for $n\leq (3\mbox{--}4)n_0$ of our interest. The latter dependence of the Dirac nucleon mass, $m_{D}^*(x)$, approximately fits that one found within the relativistic mean-field model, cf. \cite{MSTV90}.

With the free pseudo-vector $\pi NN$ coupling but for  non-relativistic nucleons without taking into account of the nucleon-nucleon correlations, calculation of the first diagram of Fig.  \ref{PionSelfenergy} is straightforward. For $N=Z$ one finds, cf. \cite{MSTV90},
\be
\Re\Pi^{0R}_P(\omega,k)\approx -\frac{2f_{\pi N}^2  m_N^* (k^2-\omega^2) p_{{\rm F}} \Phi (\omega,k)}{\pi^2}\,,\label{self-enP0}
\ee
where $\Phi (\omega,k)$ is  the Lindhard function, cf. \cite{Ericson:1988gk},
\be
\Phi (\omega,k)=\Phi_1 (\omega,k)+\Phi_1 (-\omega,-k)\,,\label{Lindh}
\ee
and  $\Phi_1 (\omega,k)$ is named the Migdal function, cf. \cite{MSTV90,Ericson:1988gk},
\be
\Phi_1 (\omega,k)=-\frac{m^{*\,2}_N}{2p_{\rm F}k^3}\left[\frac{a^2-b^2}{2}\ln  \frac{a+b}{a-b}-ab\right]\,,\nonumber
\ee
 $b\approx k p_{\rm F}/m_N^*$, $a\approx \omega +t^2/(2m_N^*)$, $t=k^2-\omega^2$.

For $\omega\gg k v_{{\rm F}}$, $v_{{\rm F}}=p_{{\rm F}}/m_N^{*}$, $N=Z$, one gets
\be \Phi(\omega\gg k v_{{\rm F}},k, T=0)\approx -\frac{\pi^2 n t}{m_N^{*2}p_{{\rm F}}\omega^2}\,,\label{Lindh-high}
\ee
 and the contribution $\Re\Pi_P(\omega,k)$ is tiny (repulsive   for $k^2>\omega^2$ and attractive otherwise, cf. \cite{Voskresensky:1993ud}).

Oppositely, in the region of  the pion frequencies $\omega\ll kv_{{\rm F}}< 4\epsilon_{{\rm F}}$,  where   $\epsilon_{{\rm F}}(n_0)\approx 40$ MeV, the Lindhard function
$\Phi(\omega,k)$   is given by \cite{Voskresensky:1978cb,Voskresensky:1982vd},
 \be
 \Re\Phi(\omega\ll kv_{{\rm F}}, k,T)\approx 1-\frac{k^2}{12p_{{\rm F}}^2}-\frac{\omega^2}{k^2v_{{\rm F}}^2}-\frac{\pi^2 T^2}{12\epsilon_{{\rm F}}^2}\,,
 \label{Phi-exp}
 \ee
 \be \Im \Phi(\omega\gg kv_{{\rm F}}, k)\approx \frac{\pi m_N^* \omega }{2kp_{\rm F}}\,.\ee
 Further assuming that  $T\leq (0.3\mbox{--}0.5)\epsilon_{{\rm F}}$   the $T$ dependence of the pion self-energy will be neglected.

References \cite{Troitsky1981,MSTV90,Voskresensky:1993ud,Voskresensky:2022gts}, as a reasonable possibility, suggested to employ the low energy theorems and found then that for $N=Z$ the residual regular contribution to the pion self-energy is
\be
\Re\Pi_{\rm reg}^{\rm off}(\omega, k)\approx C_{\rm reg}[2 k^2  + (m^2_\pi-\omega^2)] n\,.\label{reg}
\ee
$C_{\rm reg}=\Sigma/(f^2_\pi m^2_\pi)$, $\Sigma$ is the  pion-nucleon sigma
term,  $f_\pi\approx 92$ MeV is the weak pion decay constant. From the fit of the pion atom data Refs.  \cite{Troitsky1981,MSTV90,Voskresensky:1993ud} found  $C_{\rm reg}\approx 0.5/m^3_\pi\,$ that corresponds to $\Sigma\approx 30$ MeV.

To estimate a possible $NN$ correlation effect in Eq. (\ref{reg}), in this work I   conjecture the replacement
 \be C_{\rm reg}\to C_{\rm reg}^*=C_{\rm reg}\Gamma_{\rm reg}\,,\,\, \Gamma_{\rm reg}=\frac{1+c_{\rm reg}}{1+c_{\rm reg}x}\,.\label{Cregcor}
 \ee
  With $c_{\rm reg}\approx 0; 0.3; 0.5$ it is obtained $\Sigma \approx 30; 40; 45$\,MeV, respectively.

Works, cf. \cite{Migdal:1974jn,Baym:1975tm,Voskresensky:1982vd,Ericson:1988gk} used for $N=Z$ the simplest choice
\be\Pi_{\rm reg}^{\rm  off}(\omega, k)=0\,,\label{reg-tr}
\ee
satisfying the experimental result that $m^*_\pi (n\lsim n_0)\approx m_\pi$. In \cite{Voskresensky:2022gts} this choice, being motivated within the sigma model, was labeled as ``SM2,off''.
With such a choice, to appropriately reproduce the pion spectrum at $\omega\approx m_\pi$, $k\to 0$ it was used the value $f_{\pi N\Delta}\approx 1.7 m^{-1}_\pi$,  being motivated within the chiral-symmetrical model, cf. \cite{Baym:1975tm}.

One also employed  the on-mass-shell conjecture, cf. \cite{Delorm1992,Erikson1994,Kolomeitsev2003}. At such a treatment
\be \Re\Pi_{\rm reg}^{\rm on}=C_{\rm reg}(\omega^2 -m^2_\pi)n\,,\label{reg-on}
\ee
\be\Re\Pi_{\rm reg}^{\rm on}(\omega^2=m^2_\pi+k^2)=\Re\Pi_{\rm reg}^{\rm off}(\omega^2=m^2_\pi+k^2)\,.\label{reg-on1}
\ee
Reference \cite{Kolomeitsev2003} found  $C_{\rm reg}\approx 0.74/m^3_\pi\,$ that corresponds to $\Sigma\approx 45$ MeV.
The off-shell behaviors of $\Re\Pi_{\rm reg}^{\rm on}$ and $\Re\Pi_{\rm reg}^{\rm off}$ are opposite, the former quantity yields attraction at $\omega =0$,
whereas the latter one produces  repulsion.  With the parametrization (\ref{reg}) at $N=Z$ the s wave pion condensation and the  crystal-like pion condensation    do not occur up to high densities at least by the second-order phase transition.  Within the on-mass-shell conjecture the $s$ wave condensation may appear   provided the $\omega^2$ term changes  sign for $n>n^s_\pi$, cf. \cite{Voskresensky:2022gts}. If the gas parametrization  (\ref{reg-on1})  held for $n>n_0$,  the $s$ wave pion condensation at $N=Z$ could  occur in this model already for $n_c^s=1/C_{\rm reg}\leq (1.4\mbox{--} 2.5) n_0$,  as it was demonstrated in \cite{Voskresensky:2022gts}.  At $n=n_c^s$  for $\omega=k=0$ the quantity $m^{*2}_\pi (\omega =0) =m^2_\pi(1-C_{\rm reg}n_0 x)$ and $\widetilde{\omega}^2$ in Eq. (\ref{omtild}) reach zero. The sub-leading contribution to the pion self-energy in the chiral perturbation expansion  proves to be tiny,  $\delta\Pi_{\rm sl} \approx \omega^2 p_{\rm F}^4/(8\pi^4f_\pi^4)$, the sub-sub-leading term was only roughly estimated, $\delta\Pi_{\rm ssl} \approx -3n_0^2 x^2 (Dm^2_\pi+E\omega^2)/(16f^2_\pi)$ since the coefficients $D$ and $E$ were not calculated, cf. \cite{Reddy2023}. On the other hand, it is stated that the chiral perturbation theory well describes nuclear properties for $n\lsim n_0$. In order the chiral  expansion held properly, the $N$ order-expansion term should be at least two-three times smaller than that of $N-1$ order one. The condition $|\delta\Pi_{\rm ssl}/\delta\Pi_{\rm sl}|<1/2$ holds for $x\leq 1.5$  and $\omega\approx m_\pi$ only for  $|D+E|<0.2$. However for such values of $|D+E|$ the correlation effects  almost do not  change  the value of $n_c^s$ estimated in \cite{Voskresensky:2022gts} in the gas approximation. On the other hand, if $|D+E|$ were $> 0.5$, the chiral  expansion  would not be  applicable in this problem even for $n\sim n_0$. Thus, at present time one cannot  solidly constrain  the correlation parameter $c_{\rm reg}$ except that in Eq. (\ref{Cregcor}) it is connected with the value of the nucleon $\Sigma$ term.

  Note that for $\omega =0$, $k=0$  Eqs. (\ref{reg}), (\ref{reg-tr}) and (\ref{reg-on}) differ only by an upward or downward  shift of the effective pion gap (mass) $m^{*2}_\pi (\omega =0)$. A more important difference comes from the fact that (\ref{reg}) contains extra $k^2$ repulsive term. This
 difference  of Eq. (\ref{reg}) from Eqs.  (\ref{reg-tr}) and (\ref{reg-on})  can be hidden in  the pre-factor $\xi_0$ in the $\Re\Pi_\Delta$ and the value of the $g'$ parameter, being fitted to describe pion atoms at $\omega\approx m_\pi$, $k\ll m_\pi$ and the Gamov-Teller transitions for  $\omega\ll m_\pi$, $k\sim p_{\rm F}$. The given study  focuses on   description of  possibility of the $p$ wave  pion condensation at the pion energy $\omega =0$ and for $k\neq 0$ in two interpenetrating beams of nuclei in peripheral heavy-ion collisions.

\section{$NN$ interaction amplitude for $n>n_{c1}$}\label{Amplitude}
\begin{figure}\centering
\includegraphics[width=3.8cm,clip]{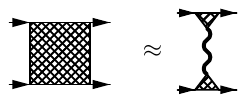}
\caption{Amplitude of $NN$ interaction in the model of in-medium pion exchange. Bold wavy line shows dressed pion Green function, hatched vertices take into account $NN$ correlations.
 }\label{NNamplitude}
\end{figure}

\begin{figure}\centering$ \boldsymbol{\dots}\parbox{4cm}{
\includegraphics[width=3.8cm,clip]{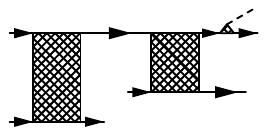}}$
\caption{Amplitude of production of a particle (shown by dashed line) after multiple $NN$ scatterings.
 }\label{Multiparticle}
\end{figure}

The amplitude, $F_{NN}$, of the $NN$ interaction contains contribution of the medium one-pion exchange (MOPE) and the contribution of the residual $NN$ interaction. In the Fermi liquid approach the latter term is parameterized with the help of the Landau-Migdal parameters. For $n\gsim n_{c1}$ with increasing density the MOPE term becomes dominant, e.g., cf. \cite{MSTV90,Voskresensky:2001fd}.
In the MOPE model   the amplitude of the $NN$ interaction, $F_{NN}[{\rm MOPE}]$, is symbolically shown in Fig. \ref{NNamplitude}. Approximately one has
\be F^R_{NN}[{\rm MOPE}]=\frac{f_{\pi N}^2 \vec{\sigma}_1\vec{k} \cdot\vec{\sigma}_2\vec{k}\,\Gamma^2(g^\prime , \omega,k,n)}{\omega^2 -m^2_\pi -\Pi^R (\omega,k,n)}\,,\ee
$\vec{\sigma}_{1,2}$ are Pauli matrices acting on the  nucleons 1 and 2, respectively.
The retarded amplitude of the $NN$ interaction enters the cross sections of all particle production processes occurring after the single and multiple  $NN$ collisions, see Fig. \ref{Multiparticle}.

The ratio of the $NN$ cross section calculated with the free one-pion exchange, FOPE, and the medium one-pion exchange for $\omega\ll kp_{\rm F}/m_N^*$, $k\sim k_0$ at $n>n_{c1}$, is as follows, e.g., cf. \cite{Voskresensky:2001fd},
\be R=\frac{\sigma[{\rm FOPE}]}{\sigma[{\rm MOPE}]}\approx \frac{\Gamma^4 (g^{\prime},n)(m^2_\pi +k_0^2)^2}{\widetilde{\omega}^4 (k_0^2,n)+\beta^2(k_0)\omega^2}\,,\label{ratio}\ee
where $\beta$ is determined in Eq. (\ref{beta}).

For $n>n_{c1}$ the squared effective pion gap $\widetilde{\omega}^2(k^2)$ gets minimum at $k=k_0\neq 0$ and $\widetilde{\omega}^2(k_0^2)\approx m^{*\,2}_\pi -\xi^2/(4\alpha_2)$, where $\xi=1-\alpha_1(n)-4\pi\alpha_h^2(n)$, decreases with increasing $n$. In assumption that  the crystal-like pion condensate occurs by the second-order phase transition, one has
$\widetilde{\omega}^2(k_0^2)\to 0$ for $n\to n^\pi_{c}$ and $R$ is increased significantly.

In case of freely interpenetrating nuclei the ratio (\ref{ratio}) changes to
\be R_{(2)}=\frac{\sigma_{(2)}[{\rm FOPE}]}{\sigma_{(2)}[{\rm MOPE}]}\approx \frac{\Gamma^4_{(2)} (g^{\prime},\theta,n)(m^2_\pi +k_0^2)^2}{\widetilde{\omega}_{(2)}^4 (k_0^2,\theta, n)+\beta_{(2)}^2\omega^2}\,,\label{ratio1}\ee
$\beta_{(2)}\simeq \gamma^2(n)\beta(n/2)/2$ for $\vec{k}\perp \vec{p}_l$ and $\beta_{(2)}\simeq \beta(n/2)$ for $\vec{k}\parallel \vec{p}_l$. Equation (\ref{ratio1}) shows strong angular dependence of the $NN$ interaction amplitude in case of nonoverlapped nucleon Fermi spheres, cf. Eqs. (\ref{Phishift}), (\ref{self-enP2}).

\end{document}